\documentclass[journal=jacsat,manuscript=article]{achemso}
\usepackage[T1]{fontenc}
\usepackage{epstopdf}
\usepackage[normalem]{ulem}
\usepackage{xcolor} 
\usepackage[colorlinks]{hyperref}

\newif\ifcomments\commentstrue

\definecolor{YRblue}{HTML}{1A56DB}

\hypersetup{
    colorlinks=true,
    linkcolor=blue, 
    urlcolor=blue, 
    citecolor=blue, 
}
\AppendGraphicsExtensions{.tif}

\usepackage[version=3]{mhchem} 
 
\author{Muhammad Hassan Shaikh}
\affiliation{Department of Physics, University of Delaware, Newark, DE, 19716, USA}
\email{mhshaikh@udel.edu}
\author{Daria Blach}
\affiliation{Molecular Foundry, Lawrence Berkeley National Laboratory, Berkeley, CA 94720, USA}
\author{Harishankar Jayakumar}
\affiliation{Molecular Foundry, Lawrence Berkeley National Laboratory, Berkeley, CA 94720, USA}

\author{Brayan Eraso Solarte}
\affiliation{Department of Physics, University of Delaware, Newark, DE, 19716, USA}
\author{Kenji Watanabe}
\affiliation{National Institute for Materials Science (NIMS), Tsukuba, 305-0047, Japan}
\author{Takashi Taniguchi}
\affiliation{National Institute for Materials Science (NIMS), Tsukuba, 305-0047, Japan}
\author{Yafei Ren}
\affiliation{Department of Physics, University of Delaware, Newark, DE, 19716, USA}
\email{yfren@udel.edu}
\author{Archana Raja}
\affiliation{Molecular Foundry, Lawrence Berkeley National Laboratory, Berkeley, CA 94720, USA}
\alsoaffiliation{Kavli Energy NanoScience Institute, University of California, Berkeley,  CA 94720, USA}

\author{Chitraleema Chakraborty}
\affiliation{Department of Physics, University of Delaware, Newark, DE, 19716, USA}
\alsoaffiliation{Quantum Science and Engineering Program, University of Delaware, Newark, DE, 19716, USA}
\alsoaffiliation{Department of Material Science and Engineering, University of Delaware, Newark, DE, 19716, USA}
\email{cchakrab@udel.edu}
\phone{+1(302) 831-3251}

\title
  {\textbf{On-demand valley control in a WSe$_2$/CrPS$_4$ heterostructure via optically programmed magnetic order}}

\begin{document}

\begin{abstract}
The ability to optically control the valley degree of freedom in two-dimensional semiconductors is critical for scalable valleytronics. While magnetic proximity effects can lift valley degeneracy, static interfaces lack the dynamic control needed for devices. Here, we demonstrate on-demand optical switching of valley symmetry breaking in a WSe$_2$/CrPS$_4$ heterostructure by exploiting light-tunable magnetic order. Using polarization- and time-resolved photoluminescence, we show that excitation below the ligand-to-metal charge transfer (LMCT) gap preserves an out-of-plane N\'eel vector in CrPS$_4$, generating a net out-of-plane spin at the interface and enabling spin-polarized charge transfer. This yields a robust polarization gap in the WSe$_2$ A-exciton decay dynamics without an external magnetic field. Excitation above the LMCT gap extinguishes the polarization gap, indicating a rotation of the N\'eel vector to in-plane. This reversible switching provides direct evidence of spin-polarized charge transfer. Furthermore, the same optical excitation enables optical control of the decay dynamics of the low-energy excitons (X$_{LE}$) by $\sim$42\%, which is potentially attributed to the modification of interfacial couplings. Our work establishes a first unique multifunctional platform where magnetic symmetry, valley polarization, and exciton relaxation are integrated and dynamically controlled by light.
\end{abstract}

 
\section{Introduction}
Two-dimensional (2D) van der Waals (vdW) heterostructures have emerged as a versatile platform for engineering quantum phenomena by design. By combining magnetic materials with optically active semiconductors, it is now possible to control spin, valley, and excitonic degrees of freedom through proximity effects \cite{zhang2020gate,beer2024proximity,shaikh_magnetic_2025,chen_layer-resolved_2025,shaikh_anomalous_2026,shaikh_optically_nodate}. A particularly compelling goal is the on-demand manipulation of valley symmetry, a prerequisite for valleytronics without external magnetic fields. While magnetic proximity can lift valley degeneracy in transition metal dichalcogenides (TMDCs) \cite{li_interfacial_2026,zhong2017van,zhao_others_2017,zhong_layer-resolved_2020}, static magnetic ordering imposes fixed operational conditions. A dynamic, optically controlled magnetic interface that can switch between distinct symmetry-breaking states would unlock programmable valley functionality.

Antiferromagnetic (AFM) materials are promising candidates for such dynamic control due to their ultrafast spin dynamics and sensitivity to optical excitation \cite{gao_photoexcitation_2023,liu_light-tunable_2020,zhang2020gate}. When interfaced with a TMDC such as WSe$_2$, these magnetic changes can translate into tunable valley polarization \cite{shaikh_magnetic_2025,chakraborty2024flatland,zhong2017van,shaikh_optically_nodate}. However, direct experimental evidence of dynamically controlled valley symmetry remains elusive.

Here, we demonstrate optical control of valley symmetry breaking in a WSe$_2$/CrPS$_4$ heterostructure by exploiting the excitation-dependent magnetism of CrPS$_4$. Using polarization- and time-resolved photoluminescence (P-TRPL), we show that excitation below the ligand-to-metal charge transfer (LMCT) gap of CrPS$_4$ preserves an out-of-plane N\'eel vector orientation, which generates a net out-of-plane spin component at the interface and enables spin-polarized charge transfer across the heterojunction. This results in a robust polarization gap in the decay dynamics of the A-exciton (X$_{A}$) in WSe$_2$. Notably, this symmetry breaking is achieved without any applied magnetic field. Excitation above the LMCT gap, however, extinguishes the polarization gap, suggesting a photo-doping-induced rotation of the N\'eel vector from out-of-plane to in-plane. This light-driven switching is reversible, providing on-demand valley control through light-induced spin-polarized charge transfer and demonstrating that the symmetry breaking can be dynamically switched on and off solely by tuning the excitation energy.

Beyond valley control, we reveal that the same optical excitation tunes the decay dynamics of low-energy excitons (X$_{LE}$) in WSe$_2$, modifying their decay time constants. By analyzing radiative and non-radiative lifetimes, we show that the non-radiative decay time decreases sharply as the temperature approaches the phase transition, while the radiative lifetime increases and saturates above this temperature. The $\sim$42\% increase in the X$_{LE}$ decay time constant under above-LMCT excitation highlights the potential of magnetic interfaces for stabilizing and controlling localized emitters by controlling their decay dynamics.

Our work establishes a multifunctional platform where magnetic symmetry, valley polarization, and exciton emission are integrated and controlled by light in a single heterostructure. These findings provide a pathway toward optically programmable valleytronics and open new avenues for probing spin-exciton physics in 2D magnetic heterostructures.

\section{Results and discussion}

\begin{figure}[!ht]
    \includegraphics[width=1\textwidth]{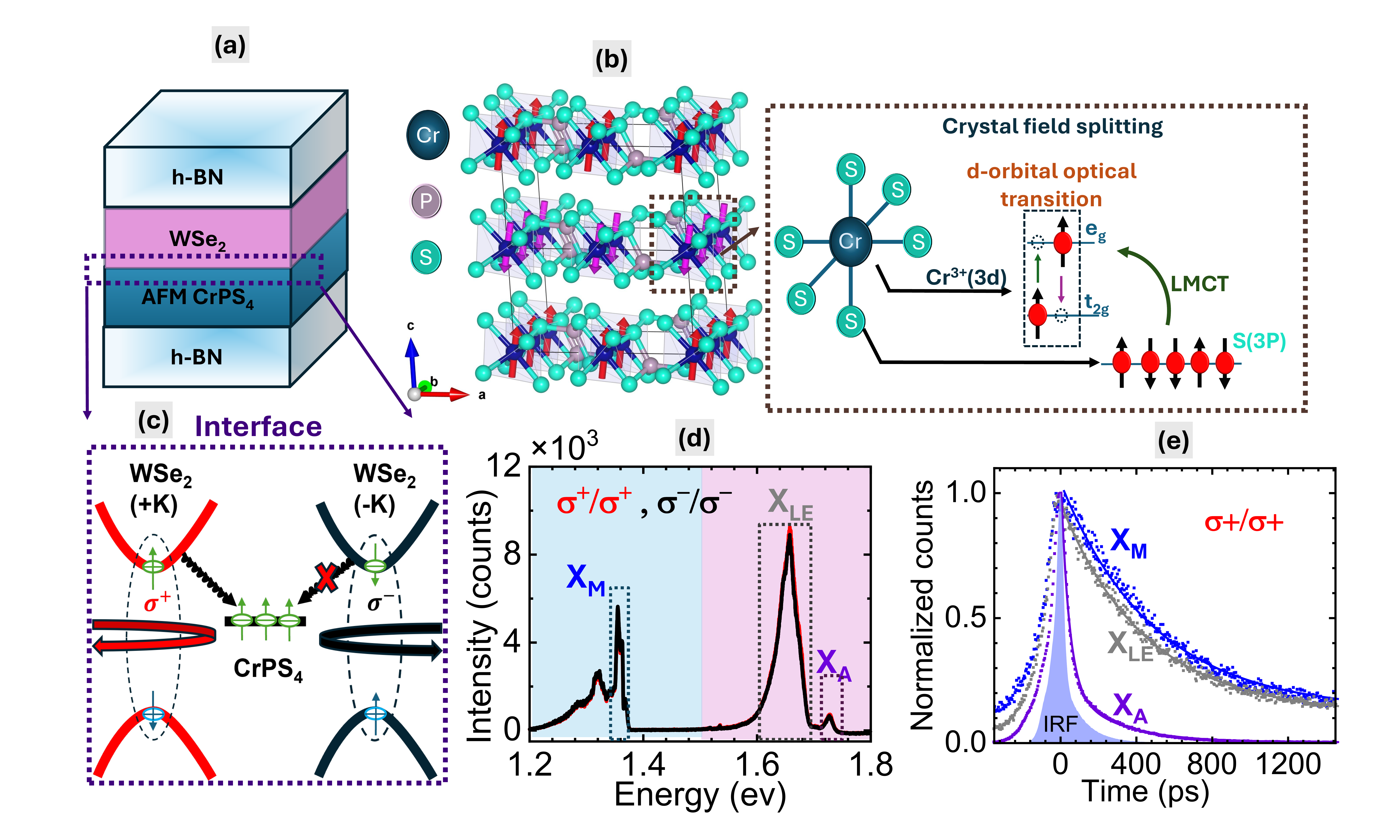}
    \caption{\textbf{Time- and polarization-resolved photoluminescence measurements of the \ce{CrPS4}/\ce{WSe2} heterostructure.}
    (a) Schematic of the encapsulated van der Waals heterostructure composed of \ce{CrPS4} and \ce{WSe2}. (b) Crystal structure of \ce{CrPS4}, where each Cr atom is octahedrally coordinated by six sulfur atoms, forming \ce{CrS6} chains along the $b$-axis. These chains are connected along the $a$-axis through phosphorus atoms. The schematic also shows two distinct optical absorption pathways in \ce{CrPS4}: (1) $d$-orbital optical transitions and (2) ligand-to-metal charge transfer (LMCT), arise from ligand-field perturbation in \ce{CrPS4}. (c) Type-II band alignment at the interface of \ce{CrPS4} and \ce{WSe2}, facilitating spin-polarized charge transfer from \ce{WSe2} to \ce{CrPS4} at the interface. (d) Time-integrated, polarization-resolved PL spectrum of the \ce{CrPS4}/\ce{WSe2} heterostructure. (e) Time-resolved PL of the heterostructure under $\sigma^+$/$\sigma^+$ configuration, with labels corresponding to the optical transitions shown in (d). The solid lines are exponential fits, while the scatter points represent the raw data.}
    \label{F1}
\end{figure}

In this work, we study the effect of changes in the magnetic structure of the vdW AFM \ce{CrPS4} on the exciton dynamics of \ce{WSe2} via the magnetic proximity effect. The device schematic used to study this effect is shown in Figure \ref{F1}a, where an AFM \ce{CrPS4} layer is placed in close proximity to an optically active semiconductor, monolayer \ce{WSe2}. The heterostructure is encapsulated by hexagonal boron nitride (h-BN). The magnetic heterostructure was prepared using a PDMS-assisted dry transfer technique \cite{shaikh_bridging_nodate}; the optical image of the heterostructure is shown in Supplementary Figure S1. CrPS\(_4\) is an A-type antiferromagnet, with ferromagnetic spin alignment within each layer and antiferromagnetic coupling between adjacent layers (Figure \ref{F1}b). Bulk CrPS\(_4\) develops long-range AFM order below approximately \(T_N\approx 37\) K. In the monolayer limit, where the interlayer antiferromagnetic coupling is absent, CrPS\(_4\) becomes a two-dimensional ferromagnet with an experimentally reported \(T_C\approx 25\) K. The crystal structure of \ce{CrPS4} is monoclinic, with each Cr atom octahedrally coordinated by six sulfur atoms, forming \ce{CrS6} chains along the $b$-axis. These chains are connected along the $a$-axis via phosphorus atoms \cite{airstable,metamagnetic,buccoliero_ligand-mediated_2026,fas_direct_2025,shaikh_magnetic_2025} (Figure \ref{F1}b). 
Recent studies show that the octahedral coordination splits the Cr $d$-orbitals into a ground state $t_{2g}$ and an excited state $e_g$ via ligand-field perturbation, which governs the optical transitions in \ce{CrPS4} \cite{lee_structural_2017,sternemann_orbital-selective_2026, buccoliero_ligand-mediated_2026}. Previous absorption studies of bulk \ce{CrPS4} show two absorption peaks: a low-energy peak near 1.7~eV (730~nm) corresponding to $d$-$d$ optical transitions, and a high-energy peak near 2.9~eV (428~nm) corresponding to the LMCT gap, with a band edge close to 2~eV (620~nm) \cite{lee_structural_2017,buccoliero_ligand-mediated_2026}. Both absorption features are illustrated schematically in Figure \ref{F1}b. To understand how the different absorption channels in AFM \ce{CrPS4} affect the exciton dynamics in \ce{WSe2}, we performed polarization-resolved spectroscopy on the \ce{CrPS4}/\ce{WSe2} magnetic heterostructure (Figure \ref{F1}a). This device architecture enables dual optical readout of magnetism from \ce{CrPS4}. The degree of circular polarization of the optical transitions in \ce{CrPS4} directly probes its bulk magnetic structure\cite{shaikh_magnetic_2025,fas_direct_2025,shaikh_optically_nodate}, while \ce{WSe2} detects the surface-layer magnetic state via type-II band alignment and spin-polarized charge transfer at the interface. This spin-polarized charge transfer allows only same-spin carriers to transfer from \ce{WSe2} to \ce{CrPS4} while blocking opposite-spin carriers at the interface, creating an enhancement in one polarization optical transition in \ce{WSe2} and producing a finite polarization contrast that can be used to study the surface-layer spin structure (Figure \ref{F1}c) \cite{shaikh_magnetic_2025,chakraborty2024flatland,chen_layer-resolved_2025,shaikh_optically_nodate}.

We excited the sample with two different laser pulse energies: one below the LMCT gap (695 nm, 1.78 eV) and one above the LMCT gap (532 nm, 2.33 eV), and performed polarization-resolved spectroscopy. Both the excitation and detection polarizations were circularly polarized light. We excited with $\sigma^+$ (or $\sigma^-$) light and detected the same polarization component ($\sigma^+$ or $\sigma^-$). The time-integrated spectrum of the magnetic heterostructure is shown in Figure \ref{F1}d, where features between 1.2 and 1.36~eV originate from $d$-$d$ optical transitions (X$_M$) in \ce{CrPS4}\cite{kim2022photoluminescence}, while those between 1.55 and 1.75~eV arise from optical transitions in monolayer \ce{WSe2}. Within the \ce{WSe2} region, the X$_{LE}$ peak near 1.655~eV is attributed to impurities, inhomogeneities, and defects; the peak near 1.725~eV corresponds to X$_A$, which is a band-to-band transition \cite{shaikh_magnetic_2025}. We then performed P-TRPL (Figure \ref{F1}e) on each optical transition window highlighted by dashed rectangles in Figure \ref{F1}d in the magnetic heterostructure. Details regarding the P-TRPL measurement can be found in the Methods section. The P-TRPL spectrum under the $\sigma^+$/$\sigma^+$ configuration is shown in Figure \ref{F1}e. The scatter points represent the raw data, and the solid lines are exponential fits. The decay dynamics of the X$_{LE}$ and X$_A$ in \ce{WSe2} were fitted using a biexponential function, as described by Equations 1 and 2. In contrast, the excitonic transitions in \ce{CrPS4} (X$_M$) were fitted using a single-exponential decay function. The P-TRPL spectrum of the magnetic heterostructure under the $\sigma^-$/$\sigma^-$ configuration is shown in Supplementary Figure S2.

\begin{align}
I(\sigma^+/\sigma^+, t) &= A_1^+ e^{-t/t_1^+} + A_2^+ e^{-t/t_2^+} \\
I(\sigma^-/\sigma^-, t) &= A_1^- e^{-t/t_1^-} + A_2^- e^{-t/t_2^-}
\end{align}

In Equations 1 and 2, $A_1^+$ and $A_2^+$ ($A_1^-$ and $A_2^-$) represent the amplitudes of the fast ($t_1^+$, $t_1^-$) and slow ($t_2^+$, $t_2^-$) decay components of the $\sigma^+$/$\sigma^+$ ($\sigma^-$/$\sigma^-$) decay curves, respectively.

\section{Optically Controlled Valley Splitting in the A-Exciton}

\begin{figure}[!ht]
    \includegraphics[width=1\textwidth]{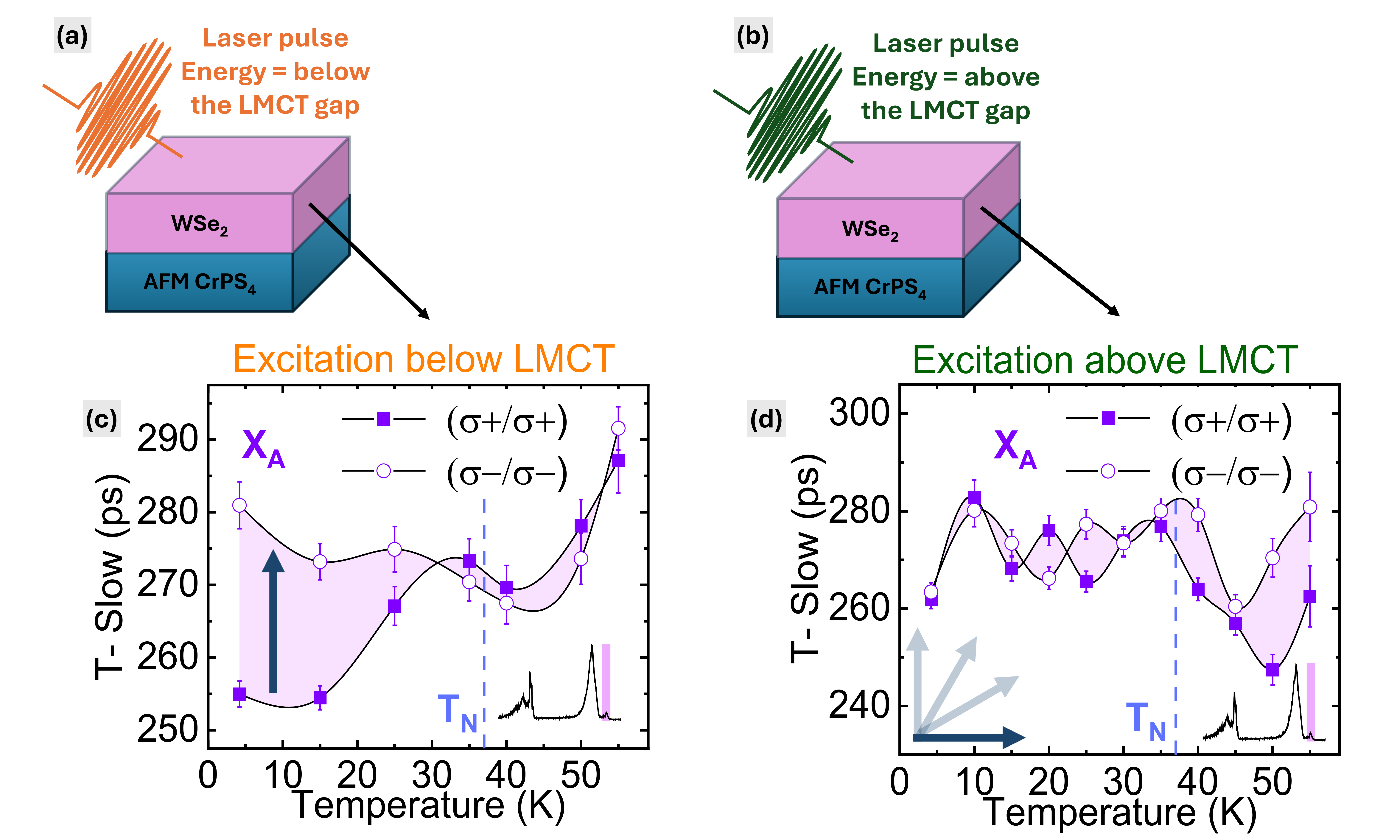}
    \caption{\textbf{Temperature dependence of the A-exciton decay time constants in the magnetic heterostructure under below- and above-LMCT excitation.}
    (a), (b) Schematic representation of the heterostructure excited with a laser pulse energy below the LMCT gap (695 nm, 1.78 eV) and above the LMCT gap (532 nm, 2.33 eV), respectively.  (c), (d) Polarization-resolved slow decay time constants of the A-exciton under below- and above-LMCT excitation, respectively. The blue arrow show the resultant N\'eel vector orientation in \ce{CrPS4}. The highlighted region in the inset spectra shows the spectral region used for TRPL measurements.}
    \label{F2}
\end{figure}

We first analyzed the temperature-dependent decay dynamics of the X$_A$ exciton under excitation with 100 picosecond (ps) laser pulses tuned both below (695 nm) and above (532 nm) the LMCT gap of \ce{CrPS4} (Figures \ref{F2}a,b). The N\'eel temperatures of \ce{CrPS4} is marked in each measurement as $T_N$ to provide insight into the temperature evolution of excitonic transitions in \ce{WSe2} due to the magnetic states of \ce{CrPS4}. When the heterostructure was excited below the LMCT gap, we observed the delay between the $\sigma^+$/$\sigma^+$ and $\sigma^-$/$\sigma^-$ polarized components of the X$_A$ TRPL (Supplementary Figure S6). This delay vanishes when the heterostructure was excited with an above-LMCT-gap laser pulse energy (Supplementary Figure S6). We also observed a robust polarization gap, defined as the difference in decay time constants between the $\sigma^+$/$\sigma^+$ and $\sigma^-$/$\sigma^-$ configurations, in both the fast and slow decay components of the A-exciton (Supplementary Figure S3 and Figure \ref{F2}c). This polarization gap vanishes in both the fast and slow decay components of the A-exciton when excited with above-LMCT-gap laser pulse energy (Supplementary Figure S3 and Figure \ref{F2}d).

The polarization gap in the decay components of X$_A$ in Figure \ref{F2}c arises due to spin-polarized charge transfer at the interface of the magnetic heterostructure. The spin-selective charge transfer at the interface (Figure \ref{F1}c) modulates the decay time constant of one polarized state ($\sigma^+$/$\sigma^+$) while the other polarized state ($\sigma^-$/$\sigma^-$) remains unchanged due to opposite spin states at the interface. This gap began to close  and exhibited a sharp increase near $T_N$, further supporting the conclusion that the observed gap arises from spin-polarized charge transfer.

The closing of the gap near the N\'eel temperature indicates that the charge transfer loses its spin polarization due to a magnetic phase transition. These results provide direct optical evidence of spin-polarized charge transfer at the \ce{CrPS4}/\ce{WSe2} interface and demonstrate that the exciton decay dynamics can serve as a sensitive probe of the magnetic phase transitions in the heterostructure.

\begin{figure}[!ht]
    \includegraphics[width=0.9\textwidth]{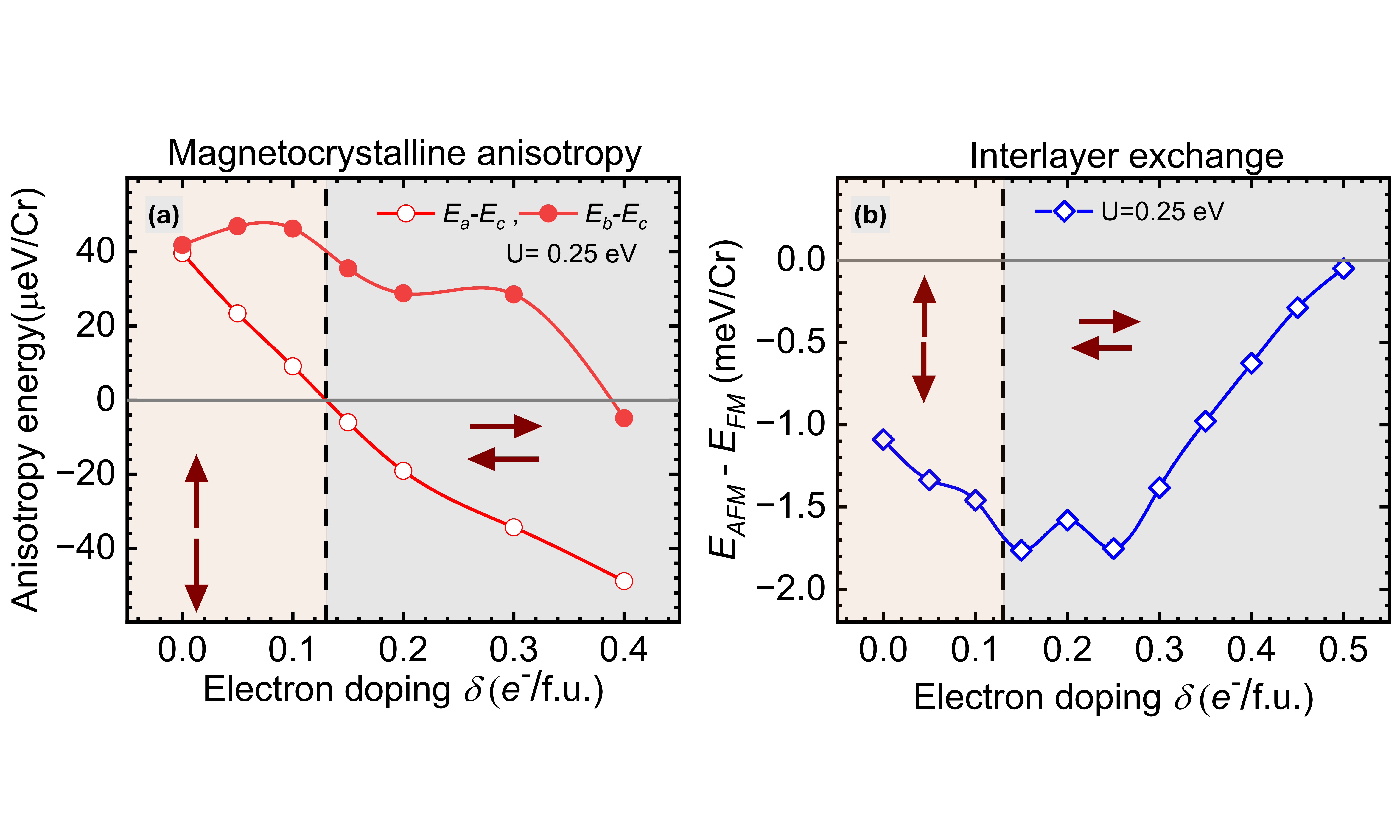}
    \vspace{-20 pt}
    \caption{Calculated magnetic anisotropy and interlayer exchange as a function of electron doping in CrPS$_4$. (a) The energy difference between the in-plane and out-of-plane spin orientations, $\Delta E_\text{MAE}$ at Hubbard $U=0.25$~eV. (b) The interlayer exchange energy, $\Delta E_\text{interlayer}$, as a function of electron doping at the same $U$. The vertical dashed line indicates the doping level at which the easy axis changes from out-of-plane to in-plane.}
    \label{fig:MAEvsU}
\end{figure}

In contrast, excitation above the LMCT gap completely suppressed this polarization gap at $\sim$4.2 K, well below the phase transition temperatures (Supplementary Figure S3 and Figure \ref{F2}d). This behavior suggests that when exciting the heterostructure above the LMCT gap, the charge transfer loses its spin polarization, resulting in the vanishing of the polarization gap in the decay time constants. This distinct behavior originates from the two disparate magnetic states of \ce{CrPS4}\cite{shaikh_optically_nodate}. Excitation below the LMCT gap preserves an out-of-plane anisotropy with the N\'eel vector oriented out of the plane, providing a net out-of-plane spin component at the interface and enabling spin-polarized charge transfer across the \ce{CrPS4}/\ce{WSe2} interface due to their type-II band alignment \cite{shaikh_magnetic_2025,chen_layer-resolved_2025,shaikh_optically_nodate}.
However, when exciting the heterostructure above the LMCT gap, a photo-doping mechanism is initiated by transferring charges from the $S(3p)$ bands to the Cr $3d$ state (Figure \ref{F1}b). This photo-induced charge transfer modifies the magnetic interactions \cite{buccoliero_ligand-mediated_2026, pandey_spinphonon_2026} and can make an in-plane spin direction lower in energy than the out-of-plane one, as predicted for CrPS$_4$ under carrier doping \cite{ruiz_controlling_2026} and found in our calculations (Figure~\ref{fig:MAEvsU}). This is also consistent with recent studies on photo-doping-induced rotation of the N\'eel vector in CrPS$_4$ \cite{shaikh_optically_nodate}. Due to this rotation of the N\'eel vector, the charge transfer loses its spin polarization, resulting in the closing of the polarization gap in the decay time constants of X$_A$. Doping-dependent changes in the magnetic exchange interaction have been predicted for monolayer \ce{CrI3} \cite{kim_exploitable_2020} and \ce{CrPS4} \cite{ruiz_controlling_2026}, and carrier-induced changes of magnetism, by optical excitation or electrostatic gating, have been reported in other materials \cite{gao_photoexcitation_2023,liu_light-tunable_2020,kim_exploitable_2020,zhang2020gate,jiang_controlling_2018,huang_electrical_2018}.

To examine whether carrier doping can reorient the N\'eel vector in CrPS$_4$, we performed DFT+$U$ calculations. We treat the excitation in the rigid electron-doping limit and calculate the magnetic anisotropy (Figure \ref{fig:MAEvsU}a) and the interlayer exchange (Figure \ref{fig:MAEvsU}b) as electrons are added to the Cr $3d$ conduction states.
In the absence of doping, the easy axis is near $c$ axis with a slight tilt toward $+a$, consistent with neutron diffraction measurements~\cite{metamagnetic,calder_magnetic_2020}. As electrons are added, the energy of spin orientation along $a$ decreases relative to $c$. Beyond a finite doping level, an in-plane direction becomes lower in energy than $c$, so that $c$ is no longer the easy axis (Figure \ref{fig:MAEvsU}a). This decrease is found for various Hubbard $U$ considered (0.25-2~eV) (Supplementary Figure S5). We focus on $U = 0.25$~eV, which gives a Kohn-Sham gap of 0.81~eV, close to the gap of 0.86~eV measured by scanning tunneling spectroscopy~\cite{sun_coupling_2025}, whereas $U = 1$-2~eV gives larger band gap of about 1~eV. With this small $U$, the interlayer coupling is still antiferromagnetic when this change of easy axis occurs as shown in Figure.~\ref{fig:MAEvsU}b. This is consistent with the persistence of antiferromagnetic order when the N\'eel vector leaves the out-of-plane direction. These calculations show that electron doping can indeed induce a change in the magnetic anisotropy without breaking the antiferromagnetic order.

\begin{figure}[!ht]
    \includegraphics[width=1\textwidth]{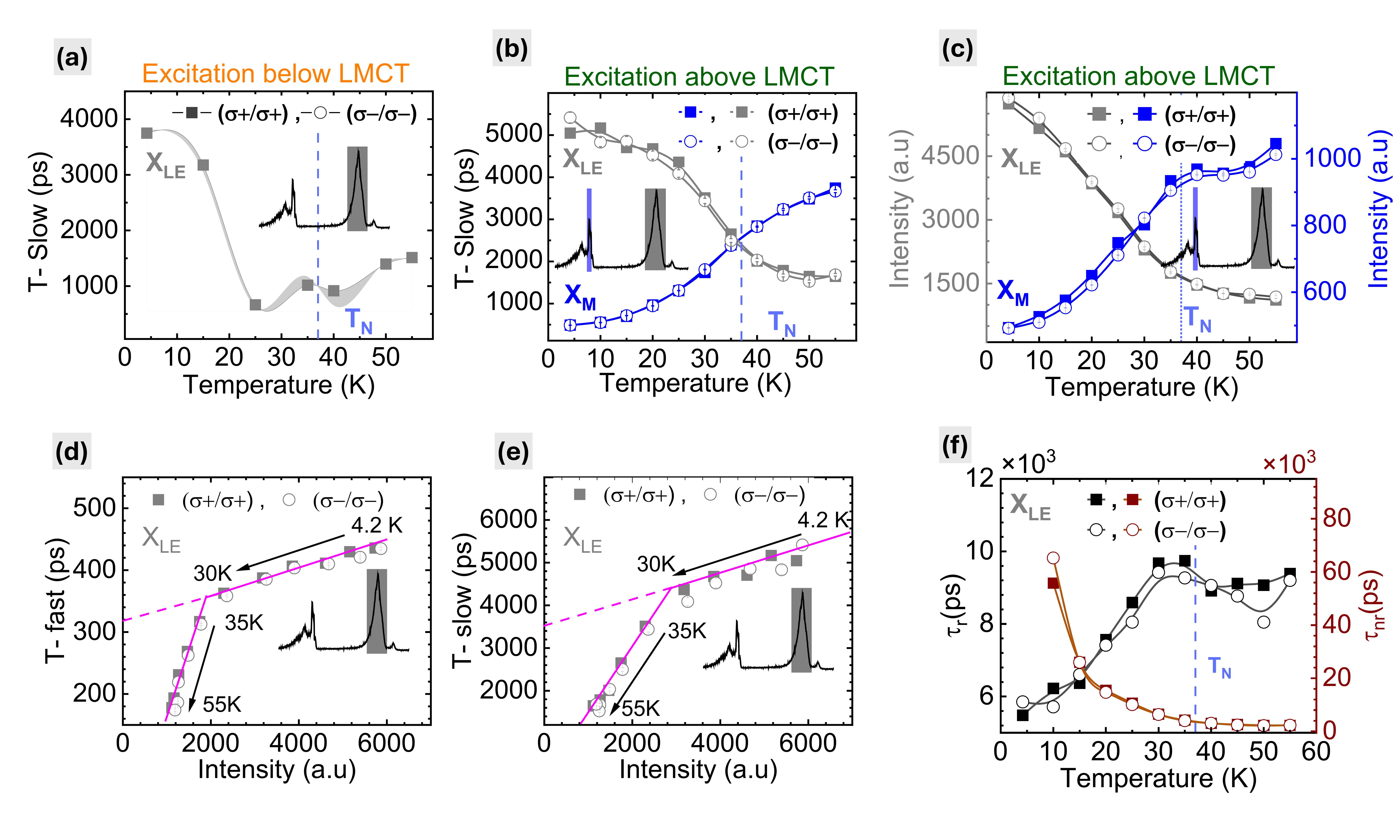}
    \caption{\textbf{Temperature-dependent dynamics of the X$_{LE}$ and X$_M$ excitons in the \ce{CrPS4}/\ce{WSe2} heterostructure.}
    (a), (b) Slow decay time constants of the LE-exciton in \ce{WSe2} under below- and above-LMCT excitation, respectively, as a function of temperature; (b) also shows the decay time constant of X$_M$ excitons in \ce{CrPS4}. The insets show the spectral regions used for the TRPL measurements. (c) Time-integrated intensity of the LE-exciton in \ce{WSe2} and the X$_M$ exciton in \ce{CrPS4} as a function of temperature under above-LMCT excitation. Both follow the same trend as their respective decay time constants. (d), (e) Slow and fast decay time constants of the LE-exciton as a function of integrated intensity. An abrupt change in the linear behavior is observed close to the phase transition temperatures. (f) Radiative and non-radiative decay time constants of the LE-exciton, calculated from the data in (d) and (e), as a function of temperature.}
    \label{F4}
\end{figure}

\section{Optically Tunable Decay Dynamics of Low-Energy Excitons}

Next, we studied the excitonic dynamics of X$_{LE}$ excitons in \ce{WSe2} as a function of temperature under both below-LMCT (695~nm) and above-LMCT (532~nm) excitation. The fast and slow decay components of the LE-exciton as a function of temperature under below-LMCT excitation are shown in Supplementary Figure S4 and Figure \ref{F4}a. We observe that when exciting the magnetic heterostructure with below-LMCT excitation, there is a sharp decrease in the decay time constants of both the slow and fast components below $T_N$, at $\sim$25~K. However, under above-LMCT excitation, the slow and fast decay components (Supplementary Figure S4 and Figure \ref{F4}b) exhibit different temperature-dependent behavior compared to the below-LMCT case. Furthermore, we observed that the decay time constant of X$_{LE}$ significantly increased under above-LMCT excitation compared to below-LMCT excitation. For example, at $\sim$4.2~K, the ($\sigma^-$/$\sigma^-$) decay component of X$_{LE}$ increases from approximately 3800~ps to 5400~ps, corresponding to a $\sim$42\% increase, when the laser pulse energy is changed from below to above the LMCT gap. Interestingly, the decay time constant of the X$_M$ exciton in \ce{CrPS4} shows an opposite trend as a function of temperature compared to the slow decay component of the X$_{LE}$ excitons (Figure \ref{F4}b), with a crossing occurring near the phase transition temperature ($T_N$), where \ce{CrPS4} transitions from the antiferromagnetic to the paramagnetic state. We also observed a change in the time-integrated intensity of X$_{LE}$ excitons in \ce{WSe2} and X$_M$ excitons in \ce{CrPS4} near this phase transition temperature, as shown in Figure \ref{F4}c.

To further understand the observed behavior of X$_{LE}$ due to the change in the magnetic ground state of \ce{CrPS4} arising from the photo-doping mechanism, we investigated the relationship between the decay components and the time-integrated PL intensity of X$_{LE}$ excitons. We plotted the fast and slow decay components of the X$_{LE}$ excitons as a function of temperature and peak intensity under above-LMCT excitation (Figures \ref{F4}d,e), respectively. For both components, we observed a linear decrease in the decay time constants and intensity with increasing temperature. However, a sharp change in slope occurs near the phase transition temperature of \ce{CrPS4}.

The defect-band transition in \ce{WSe2} is expected to decrease linearly with peak intensity due to thermally activated non-radiative depopulation channels \cite{lopion2020temperature}. To further elucidate our understanding of this sharp change in slope, we expanded our analysis by calculating the radiative and non-radiative decay times from the fast and slow decay components of X$_{LE}$ excitons as a function of temperature. The radiative lifetime is calculated following Li et al. \cite{li2020temperature}:

\begin{equation}
\tau_r(T) = \frac{\tau(T)}{I_\text{PL}(T)/I_\text{PL}(4.2\,\text{K})},
\end{equation}

where $\tau_r(T)$ is the radiative lifetime, $I_\text{PL}(T)/I_\text{PL}(4.2\,\text{K})$ is the peak intensity of the time-integrated PL as a function of temperature normalized by the peak intensity at 4.2~K, and $\tau(T)$ is the total decay time constant, given by:

\begin{equation}
\tau(T) = t_{Fast}(T) + t_{Slow}(T).
\end{equation}

The non-radiative decay time constant is obtained from the relation:

\begin{equation}
\frac{1}{\tau(T)} = \frac{1}{\tau_r(T)} + \frac{1}{\tau_\text{nr}(T)},
\end{equation}

which yields:

\begin{equation}
\tau_\text{nr}(T) = \frac{\tau(T) \cdot \tau_r(T)}{\tau_r(T) - \tau(T)}.
\end{equation}

The resulting radiative and non-radiative decay time constants for the X$_{LE}$ excitons are shown in Figure \ref{F4}f. We observe a clear decrease in the non-radiative decay time as the temperature approaches the phase transition temperature of \ce{CrPS4}, while the radiative decay time increases and saturates above this temperature.
The observed behavior of X$_{LE}$ excitons can be explained by two possible mechanisms. First, the above-LMCT excitation-induced rotation of the N\'eel vector modifies the interfacial symmetry of the \ce{CrPS4}/\ce{WSe2} heterostructure. This modification influences the phonon modes in \ce{WSe2} via magnon-phonon interaction, as explained by polarization-resolved magneto-Raman measurements of the \ce{CrPS4}/\ce{WSe2} heterostructure \cite{chen_layer-resolved_2025}, and hence modifies the X$_{LE}$ exciton dynamics. Second, the rotation of the N\'eel vector from out-of-plane to in-plane modifies the exchange magnetic field experienced by the localized X$_{LE}$ emitters, which modifies the exciton lifetimes \cite{xiang_magnetic_2025}. Due to the localized nature of X$_{LE}$ excitons, they follow the behavior of X$_M$ due to magnon-phonon coupling \cite{chen_layer-resolved_2025}. The modification of excitonic behavior in \ce{WSe2} due to phonons in \ce{CrPS4} has been observed in recent linear polarization studies \cite{chen_layer-resolved_2025}.

While exciton-phonon effects on decay time constants in A-type antiferromagnets have been previously reported in the van der Waals magnet chromium sulfur bromide (CrSBr) \cite{lin2024strong}, our findings establish excitation photon energy as a means of actively controlling the decay dynamics of localized excitons in magnetic heterostructures. In addition, we also demonstrate that we can control the decay dynamics and modify exciton-phonon coupling in localized excitons of \ce{WSe2} by controlling the interfacial symmetry by tuning the excitation energy. Furthermore, we show that the emitter decay dynamics can be controlled by $\sim$42\% through laser pulse energy tuning, without the need for an external magnetic field. This demonstrates that emitter lifetimes can be controlled when coupled to a magnetic substrate, providing a versatile platform for both valleytronic and quantum photonic applications. These findings pave the way toward high-temperature localized emitters integrated with magnetic substrates for advanced quantum devices.

\section{Conclusion and Outlook}

In conclusion, our work establishes optical control of valley symmetry breaking in a magnetic van der Waals heterostructure as a direct consequence of light-tunable magnetic proximity interactions. By simply varying the excitation energy across the LMCT gap of \ce{CrPS4}, we demonstrate deterministic switching between two distinct magnetic interfacial states. Excitation below the LMCT gap preserves an out-of-plane N\'eel vector orientation, generating a net out-of-plane spin component at the interface. This enables spin-polarized charge transfer across the \ce{CrPS4}/\ce{WSe2} interface, resulting in a robust polarization gap in the decay dynamics of the A-exciton in proximal monolayer \ce{WSe2}, observed without any external magnetic field. The temperature dependence of this polarization gap provides direct optical evidence of spin-polarized charge transfer and demonstrates that exciton decay dynamics can serve as a sensitive probe of magnetic phase transitions.

Excitation above the LMCT gap extinguishes the polarization gap, indicating a photo-doping-induced rotation of the N\'eel vector from out-of-plane to in-plane. This switching is reversible and provides an optical pathway to control the valley degree of freedom. The observed polarization gap directly signals the breaking of time-reversal symmetry in \ce{WSe2} without an applied magnetic field, a fundamental requirement for valleytronics and valley-based spintronic devices.

Furthermore, we reveal that the same optical knob modulates exciton-phonon coupling for localized low-energy excitons (X$_{LE}$) in \ce{WSe2}, enabling on-demand tuning of their decay dynamics. The $\sim$42\% increase in the X$_{LE}$ decay time constants under above-LMCT excitation, along with the sharp changes in radiative and non-radiative recombination near the phase transition temperature, underscores the potential of magnetic interfaces for stabilizing and controlling localized excitons. Our work demonstrates that emitter lifetimes can be controlled when coupled to a magnetic substrate, providing a versatile platform for both valleytronic and quantum photonic applications.

Our findings introduce a versatile platform for optically programmable valleytronics, where magnetic symmetry, valley polarization, and exciton-phonon interactions are integrated and controlled within a single heterostructure. This work moves beyond static magnetic proximity effects, demonstrating dynamic, light-driven manipulation of valley physics, a critical step toward scalable, ultrafast valley-optoelectronic and quantum photonic devices. These findings pave the way toward high-temperature localized emitters integrated with magnetic substrates for advanced quantum devices.


\section{Methods}

\subsection{Heterostructure preparation}

Bulk crystals of \ce{CrPS4}, \ce{WSe2}, and hexagonal boron nitride (h-BN) were mechanically exfoliated using low-residue tape. The cleaved flakes were further thinned by repeated exfoliation and subsequently transferred to polydimethylsiloxane (PDMS) stamps. Thin \ce{CrPS4} and h-BN flakes were identified by their optical contrast using an optical microscope, while monolayer \ce{WSe2} flakes were identified by optical contrast and further confirmed by photoluminescence (PL) spectroscopy. The heterostructure was assembled by sequentially stacking the flakes onto a Si/SiO$_2$ substrate using a PDMS-based dry transfer technique \cite{shaikh_bridging_nodate}.

\subsection{Polarization- and time-resolved photoluminescence (P-TRPL) measurements}

P-TRPL measurements were performed at Molecular Foundry, Lawrence Berkeley National Laboratory. Montana Cryostation s200 closed-cycle helium cryostat was used, which is capable of controlling the temperature from 4.2 K to room temperature. The samples were excited using a supercontinuum laser with a $\sim$100 ps pulse width and a 38.1 MHz repetition rate. The experiments were conducted under two excitation laser wavelengths: 695 nm (below the LMCT gap) and 532 nm (above the LMCT gap), using a 100$\times$ objective with a numerical aperture (NA) of 0.95. The spot sizes were $\sim$892 nm for the 695 nm laser pulse wavelength and $\sim$683 nm for the 532 nm laser pulse wavelength. The pulse energy was approximately 0.66 pJ. The average power was maintained below $\sim$25 $\mu$W to exclude laser-induced heating effects. The photon flux was calculated to be 8.72 $\times$ 10$^{13}$ photons s$^{-1}$ (1.40 $\times$ 10$^{22}$ photons cm$^{-2}$ s$^{-1}$) for 695 nm and 6.67 $\times$ 10$^{13}$ photons s$^{-1}$ (1.82 $\times$ 10$^{22}$ photons cm$^{-2}$ s$^{-1}$) for 532 nm. Circular polarization was generated and detected using a combination of a linear polarizer and a quarter-wave plate. PL spectra were collected using a spectrometer with a dispersion grating and a Peltier-cooled charge-coupled device (CCD) camera, while lifetimes were measured using an avalanche photodiode (APD) with a nominal temporal resolution of 30 ps.

\subsection{DFT calculations}

First-principles calculations were carried out with VASP 6.6.0~\cite{kresse_efficient_1996,kresse_efficiency_1996} using the PBE exchange-correlation functional~\cite{perdew_generalized_1996} and projector-augmented-wave potentials~\cite{blochl_projector_1994,kresse_ultrasoft_1999}, with the on-site Coulomb interaction on Cr treated in the rotationally invariant DFT+$U$ scheme~\cite{dudarev_electron-energy-loss_1998} and Grimme D2 dispersion corrections~\cite{grimme_semiempirical_2006}.
The plane-wave cutoff was 450~eV. All calculations use a 24-atom cell containing four Cr atoms, in which the monoclinic $c$ axis is doubled due to the A-type antiferromagnetic ground state. 
Lattice parameters are taken as $a = 10.89$~\AA, $b = 7.29$~\AA, $c = 6.13$~\AA, $\beta = 92^\circ$~\cite{ruiz_controlling_2026} and the internal coordinates relaxed at $U = 0.25$~eV. 
Photo-excitation was modelled in the rigid-doping limit. Electrons were added to the cell by changing the total electron count with a compensating uniform background, at fixed lattice parameters and internal coordinates, i.e.\ neglecting lattice relaxation in the photo-doped state. Brillouin-zone integration used $\Gamma$-centred meshes with Gaussian smearing ($\sigma = 0.05$~eV). A $6\times6\times3$ mesh was used for the undoped state and for all magnetic-anisotropy calculations. 
The interlayer exchange in the doped, metallic state is far more sensitive and required a $12\times12\times6$ mesh. Magnetic anisotropy energies were obtained with the magnetic force theorem: a collinear self-consistent charge density was computed once for each doping level, and the spin-orbit energy was then evaluated non-self-consistently for each spin direction.

\subsection{Acknowledgement}
This research was partially supported by the National Science Foundation (NSF) through the University of Delaware (UD) Materials Research Science and Engineering Center (MRSEC) Seed Award program (DMR-2011824). M.H.S. acknowledges partial support from NSF Award OIA-2217786 and NSF CAREER Award number 2543219. B.E.S. and Y.R. acknowledge support from the University of Delaware Research Foundation Strategic Initiative. Authors acknowledge work at the Molecular Foundry, which was supported by the Office of Science, Office of Basic Energy Sciences, of the U.S. Department of Energy under Contract No. DE-AC02-05CH11231, under user proposal numbers MFP-08622, MFP-09458, and MFP-10203. 
This work used Anvil at Purdue University through allocation
PHY250304 from the ACCESS program, which is supported by
U.S. NSF grants \#2138259, \#2138286, \#2138307, \#2137603, and \#2138296. Anvil is supported by the NSF under Grant No.~2005632.
The authors thank Xi Wang from the Department of Materials Science and Engineering, University of Delaware, for providing access to their transfer station and room-temperature PL and Raman measurement setup. The authors thank Yi Ji from the Department of Physics and Astronomy, University of Delaware, for providing access to their optical microscope.

\bibliography{acs-achemso.bib}

\end{document}